\documentclass[twocolumn,showpacs,preprintnumbers,amsmath,amssymb]{revtex4}
\usepackage{graphicx}
\usepackage{dcolumn}
\usepackage{bm}

\begin{document}

\title{Variation of elastic energy
shows reliable signal of upcoming catastrophic failure}
\author{Srutarshi Pradhan\email{srutarshi.pradhan@ntnu.no}}
\affiliation{PoreLab, Department of Physics, Norwegian University of
Science and Technology, NO--7491 Trondheim, Norway.}
\author{Jonas T. Kjellstadli\email{jonas.t.kjellstadli@ntnu.no}}
\affiliation{PoreLab, Department of Physics, Norwegian University of
Science and Technology, NO--7491 Trondheim, Norway.}
\author{Alex Hansen\email{alex.hansen@ntnu.no}}
\affiliation{PoreLab, Department of Physics, Norwegian University of
Science and Technology, NO--7491 Trondheim, Norway.}

\date{\today {}}

\begin{abstract}
We consider the Equal-Load-Sharing Fiber Bundle Model  
as a model for composite materials under stress and derive elastic energy and 
damage energy as a 
function of strain. With gradual increase of stress (or strain) the bundle 
approaches a catastrophic failure point where 
the elastic energy is always larger than the damage energy. 
We observe that elastic energy 
has a maximum that appears after the catastrophic failure point is passed, 
i.e., in the unstable phase of the system.
However, the slope of elastic energy 
{\it vs.\/} strain curve has a maximum which always appears before the 
catastrophic failure point and therefore this can be used as a 
reliable 
signal of upcoming catastrophic failure. We study this behavior analytically 
for power-law type and Weibull type distributions of fiber thresholds and 
compare the results 
with numerical simulations on a single bundle with large number of fibers.        
\end{abstract}
\maketitle
\section{Introduction}
Accurate prediction of upcoming catastrophic failure events has important and 
far-reaching consequences. It is a
central problem in material science in connection with the durability of 
composite 
materials under
external stress  \cite{k96,hhlp08,bb11,a15,b15}. The same problem exists at a 
large scale (field-scale) 
associated with mine and cave collapses, landslides, snow avalanches
 and 
the onset of earthquakes due to
plate movements \cite{cb97,brc15}. 
In medical science, 
understanding fracturing of human bones exposed to 
a sudden stress is an important research area \cite{vm91}. 
These phenomena belong to the class of phenomena called stress-induced 
fracturing, where
initially micro-fractures are produced here and there in the system and at
some point, due to gradual stress increase, a major fracture develops through 
coalescence of 
micro-fractures and the whole system collapses (catastrophic event).
Such stress-induced failures occur also in very different domains --- 
for example, in breakdown of social relationships and mental health \cite{e84,rgrcg96}. 

The central question is --- when does the
catastrophic failure occur? Is there any prior signature that can tell us 
whether catastrophic failure is imminent? The
inherent heterogeneities of the systems and the stress redistribution 
mechanisms 
(inhomogeneous in most 
cases) make things complicated and a concrete theory of the prediction schemes, even in model
systems, is still lacking. 

In this article, we address this problem (prediction of catastrophic events) in 
the Fiber
Bundle Model (FBM) which has been used as a standard model \cite{p26,d45,phc10,hhp15} for fracturing in 
composite materials under external stress. We will show theoretically that 
 in the 
Equal-Load-Sharing (ELS) model: 1) At the catastrophic failure point, the 
elastic energy is always larger than the damage energy. 2) The elastic 
energy variation
shows a distinct peak before the catastrophic failure point and 
this is a 
universal feature, i.e., it
does not depend on the threshold distribution of the elements in the system. 
3) The energy release
during final catastrophic event is much bigger than the elastic energy stored 
in the system at the
failure point. Our numerical results show perfect agreement with the 
theoretical estimates.

We organize our article as follows: After the brief introduction (section I), 
we define the elastic energy and the damage energy in the Fiber
Bundle Model in section II. In sections III and IV we calculate the elastic and 
damage energies of the model
in terms of strain or extension. In several subsections of sections III and IV 
we 
explore the theoretical
calculations for power-law type and Weibull type distribution of fiber 
thresholds. Simulation results are
presented and numerical results are compared with the theoretical estimates in 
these sections. We present a general analysis of elastic energy variations 
and existence of an elastic-energy maximum in section V. In
section VI we identify the warning sign of catatrophic failure by locating the 
inflection point.  Here, in 
addition to uniform and Weibull distributions,  we 
choose  a mixed threshold distribution 
and present the 
numerical results, based on
Monte Carlo simulation, to confirm the universality of the behavior in the ELS 
models. Finally, we keep some
discussions in section VII.
\section{The Fiber Bundle Model}
\begin{figure}[t]
\begin{center}
\includegraphics[width=6cm]{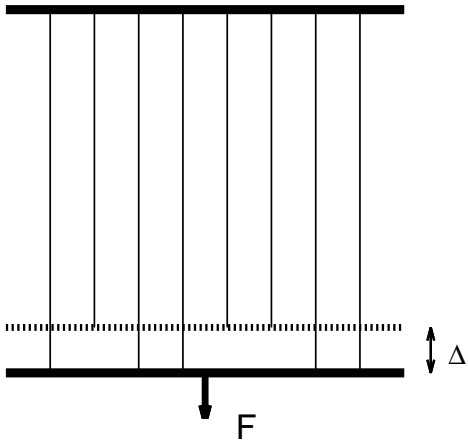}
\vskip .2in
\caption {\label{fig:model}
The fiber bundle model.}
\end{center}
\end{figure}

The fiber bundle model consists
of $N$ parallel fibers placed between two solid clamps (figure \ref{fig:model}).  
Each fiber responds
linearly with a force $f$ to a stretch or extension $\Delta$,
\begin{equation}
\label{eq1}
f=\kappa \Delta\;,
\end{equation}
where $\kappa$ is the spring constant.  $\kappa$ is the same for all fibers. 
Each fiber has a  threshold $x$ assigned to it.  If the stretch $\Delta$ 
exceeds
this threshold, the fiber fails irreversibly.  When the clamps are stiff, 
load will be redistributed equally on the surviving fibers and this is called 
the equal-load-sharing (ELS) scheme. Throughout this article we work with ELS
 models only.  

The fiber thresholds
are drawn from a probability density $p(x)$. The corresponding cumulative probability
is 
\begin{equation}
\label{eq2}
P(x)=\int_0^x\ dx' p(x')\;.
\end{equation}

When the fiber bundle is loaded, the fibers fail according to their thresholds, the weaker
before the stronger.  Suppose that $n$ fibers have failed.  
At a stretch $\Delta$, the fiber bundle carries a force
\begin{equation}
\label{eq3}
F=\kappa(N-n)\Delta=N\kappa (1-d)\Delta\;,
\end{equation}
where we have defined the {\it damage\/}
\begin{equation}
\label{eq4}
d=\frac{n}{N}\;.
\end{equation}
When $N$ is large enough, $d$ may be treated as a continuous parameter.

We will now assume that the stretch $\Delta$ is our control parameter. We can 
construct the energy budget according to continuous damage mechanics \cite{k96,phr18}.  Clearly, when we stretch the bundle with external force, work is done 
on the system. At a stretch 
$\Delta$ and damage $d$, the elastic energy stored by the surviving fibers is
\begin{equation}
\label{eq6}
E^{e}(\Delta,d)=\frac{N\kappa}{2}\ \Delta^2\left(1-d\right)\;.
\end{equation}
The damage energy of the failed fibers is given by
\begin{equation}
\label{eq7}
E^d(d)=\frac{N\kappa}{2}\ \int_0^d\ d\delta \left[P^{-1}(\delta)\right]^2\;.
\end{equation}
The total energy at stretch $\Delta$ and damage $d$ is then  
\begin{equation}
\label{eq8}
E(\Delta,d)=E^{e}(\Delta,d)+E^{d}(d).
\end{equation}

\section{Elastic energy and damage energy at the failure point}
We are going to analyze the energy relations when the bundle is  
in equilibrium. We know that there is a certain value, $\Delta=\Delta_c$, 
 beyond which catastrophic failure occurs and the system collapses 
completely. We are particularly interested in what happens at the failure 
point. Is there a universal relation between elastic energy and damage energy at the 
failure point?
 
When $N$ is large, we can reframe Eqns. (\ref{eq6},\ref{eq7}) and 
express the energies in terms of 
external stretch (or extension) $\Delta$ as
\begin{equation}
\label{eq6b}
E^{e}(\Delta)=\frac{N\kappa}{2}\ \Delta^2\left(1-P(\Delta)\right)\;.
\end{equation}
and
\begin{equation}
\label{eq7b}
E^d(\Delta)=\frac{N\kappa}{2}\ \int_0^{\Delta}\ dx \left[p(x)x^2\right]\;.
\end{equation}
The force on the bundle at a stretch $\Delta$ can be written as 
\begin{equation}
\label{eq3b}
F=\kappa(N-n)\Delta=N\kappa(1-P(\Delta))\Delta.
\end{equation}
The force must have a maximum at the failure point $\Delta_c$, therefore 
setting $dF(\Delta)/d\Delta =0$ we get
\begin{equation}
\label{eq13}
1-\Delta_c p(\Delta_c)-P(\Delta_c)=0.
\end{equation}
\subsection{Uniform threshold distribution}
We start with the simplest threshold distribution: the uniform distribution, 
which is well-known in fiber bundle research \cite{hhp15}. 
For a uniform fiber threshold distribution within the range $(0,1)$,
$p(x)=1$ and $P(x)=x$. 
Therefore we get, from Eqn. (\ref{eq13}),  
\begin{equation}
\label{eq14}
\Delta_c=\frac{1}{2}.
\end{equation}

Now putting $\Delta_c=1/2$ in Eqns. (\ref{eq6b}, \ref{eq7b}), we get 
\begin{equation}
\label{eq6c}
E^{e}(\Delta_c)=\frac{N\kappa}{16}\;,
\end{equation}
and
\begin{equation}
\label{eq7d}
E^d(\Delta_c)=\frac{N\kappa}{2}\ \int_0^{1/2}\ dx \left[x^2\right]=\frac{N\kappa}{48}\;.
\end{equation}
Therefore, the ratio between damage energy and elastic energy at the failure 
point ($\Delta_c$) is
\begin{equation}
\label{eq15}
\frac{E^d(\Delta_c)}{E^e(\Delta_c)}=\frac{1}{3} \;.
\end{equation}
 
\subsection{Power-law type threshold distribution}
Now we move to a general power law type fiber threshold distributions within 
the range $(0,1)$,
\begin{equation}
\label{eq-full}
p(x)=(1+\alpha)x^{\alpha}.
\end{equation}
The cumulative distribution takes the form
\begin{equation}
\label{eq-cumm}
P(x)= \int_0^x p(y)dy =x^{1+\alpha}.
\end{equation}

We insert the expressions for  $p(x)$ and $P(x)$ into 
 Eqn. (\ref{eq13})
and find the critical extention 
\begin{equation}
\label{eq-delta_c}
\Delta_c =\left(\frac{1}{2+\alpha}\right)^\frac{1}{1+\alpha}.
\end{equation}

\begin{figure}[t]
\begin{center}
\includegraphics[width=8cm]{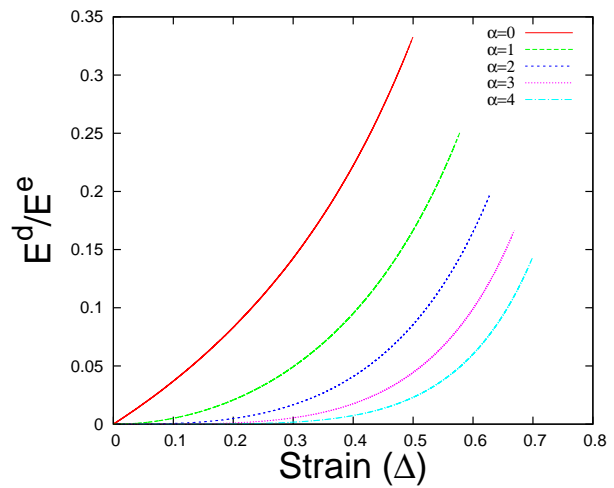}
\caption {\label{fig:ratio-vs-strain}
 Ratio between damage energy and elastic energy {\it vs.\/} extention $\Delta$. Single bundle
 with $N=10^{7}$ fibers. Different curves are for different power law 
exponents of the  fiber strength distribution. }
\end{center}
\end{figure}

\begin{figure}[t]
\begin{center}
\includegraphics[width=8cm]{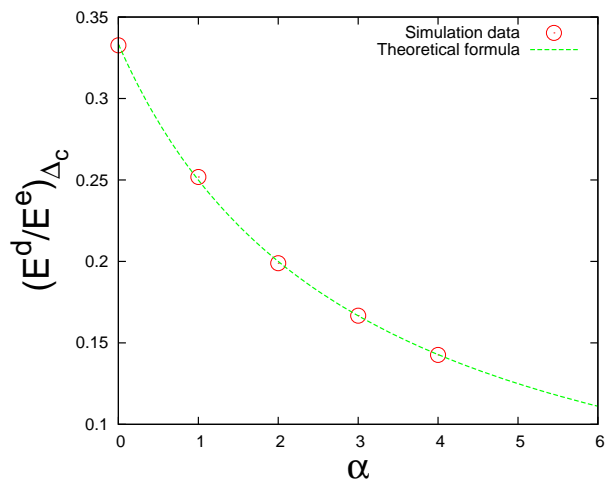}
\caption {\label{fig:ratio-vs-alpha}
 Ratio between damage energy and elastic energy at the failure point $\Delta_c$ 
{\it vs.\/} 
power law exponent $\alpha$. Single bundle
 with $N=10^{7}$ fibers. Dashed line is the theoretical estimate Eqn. (\ref{eq-energy_ratio}). }
\end{center}
\end{figure}

We can calculate the elastic energy and damage energy at the failure point 
$\Delta_c$: 
\begin{eqnarray}
\label{eq6c}
E^{e}(\Delta_c)&=&\frac{N\kappa}{2}\ \Delta_c^2\left(1-P(\Delta_c)\right)\nonumber\\ 
&=&\frac{N\kappa}{2}\ \Delta_c^2\left(1-\Delta_c^{1+\alpha}\right),\;
\end{eqnarray}
and
\begin{eqnarray}
\label{eq7c}
E^d(\Delta_c)&=&\frac{N\kappa}{2}\ \int_0^{\Delta_c}\ dx \left[p(x)x^2\right]\nonumber\\
&=&\frac{N\kappa}{2}\frac{1+\alpha}{3+\alpha} \Delta_c^{3+\alpha}\;.
\end{eqnarray}
Plugging in the value of $\Delta_c$ (Eqn. \ref{eq-delta_c}) into the above 
equations for elastic energy and damage energy we end up with the following 
relation:
    
\begin{equation}
\label{eq-energy_ratio}
\frac{E^d(\Delta_c)}{E^e(\Delta_c)}=\frac{1}{3+\alpha} \;.
\end{equation}
Clearly, the ratio depends on the power factor $\alpha$ (Figures \ref{fig:ratio-vs-strain} and \ref{fig:ratio-vs-alpha}). 
When $\alpha=0$, the threshold distribution reduces to a uniform distribution 
and we immediately go back to Eqn. \ref{eq15}. 

\subsection{Energy balance}
It is easy to show that the work done on the system 
up to the failure point $\Delta_c$ is equal to the sum of the energies 
$E^e$ and $E^d$.
The total work done on the system can be calculated as
\begin{equation}
\label{eq-work1}
W(\Delta_c)= \int_0^{\Delta_c}\ d\Delta F(\Delta)\;.
\end{equation}
Inserting the expression for $F(\Delta)$ into the integral we get, for a 
general power law type distribution, 
\begin{equation}
\label{eq-work2}
W(\Delta_c)= Nk\Delta_c^2\left[\frac{1}{2}-\frac{\Delta_c^{1+\alpha}}{3+\alpha}\right]\;,
\end{equation}
which is the total of elastic energy and damage energy, $W=E^e+E^d$ 
(see Eqns. \ref{eq6c}, \ref{eq7c}). In fact, this is the energy conservation 
relation in thermodynamic sense. 

\subsection{Energy release during the final catastrophic avalanche}

It is known that when the extension exceeds the critical value 
$\Delta_c$, the 
whole bundle collapses via a single avalanche called the final or 
catastrophic avalanche \cite{hhp15}. Can we calculate how much energy will be released in
this final avalanche?
It must be equal to the total damage energy of the fibers between threshold 
values $\Delta_c$ and the upper cutoff level of the 
fiber thresholds for the distribution in question.

We calculate the damage energy of the final avalanche for power-law type distributions as
\begin{eqnarray}
\label{eq7d}
E^d_{final}&=&\frac{N\kappa}{2}\ \int_{\Delta_c}^{1}\ d\delta \left[p(\delta)\delta^2\right]\nonumber\\
&=&\frac{N\kappa}{2}\frac{(1+\alpha)}{(3+\alpha)}\left(1- \Delta_c^{3+\alpha}\right)\;.
\end{eqnarray}
It is important to find out whether the damage energy for the catastrophic 
avalanche has a universal relation with the elastic or damage energies at the 
failure 
point. As already mentioned, the bundle has stable (equilibrium) states up 
to $\Delta \leq \Delta_c$. Therefore, if we correlate the final avalanche 
energy with $E^e$ or  $E^d$ values at $\Delta_c$,  
we can predict the catastrophic power of 
the final avalanche.

\begin{figure}[t]
\begin{center}
\includegraphics[width=8cm]{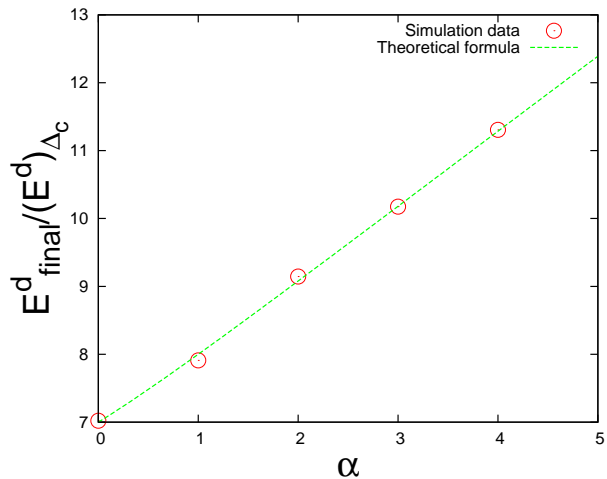}
\caption {\label{fig:final-d-vs-alpha}
 Ratio between damage energy of final catastrophic avalanche  and 
damage energy at the failure point $\Delta_c$ {\it vs.\/} 
power law exponent $\alpha$. Single bundle
 with $N=10^{7}$ fibers. Solid line is the theoretical estimate Eqn. (\ref{eq-relation1}). }
\end{center}
\end{figure}

\begin{figure}[t]
\begin{center}
\includegraphics[width=8cm]{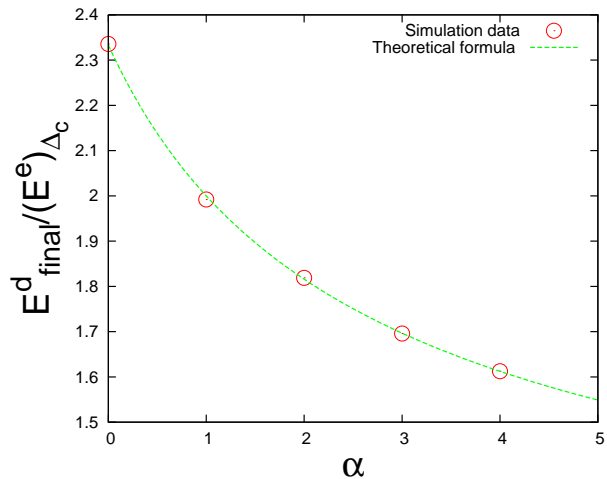}
\caption {\label{fig:final-e-vs-alpha}
 Ratio between damage energy of final catastrophic avalanche  and elastic energy at the failure point $\Delta_c$ {\it vs.\/} 
power law exponent $\alpha$. Single bundle
 with $N=10^{7}$ fibers. The dashed line is the theoretical estimate Eqn. (\ref{eq-relation2}). }
\end{center}
\end{figure}

Comparing the expressions for $E^e_{\Delta_c}$, $E^d_{\Delta_c}$ and 
$E^d_{final}$ we can write the following relation:

\begin{equation}
\label{eq-relation1}
\frac{E^d_{final}}{E^d_{\Delta_c}}=\left[(2+\alpha)^\frac{3+\alpha}{1+\alpha} -1\right]\;.
\end{equation}
As  $E^e_{\Delta_c}=(3+\alpha)E^d_{\Delta_c}$, we can easily get the other relation:
\begin{equation}
\label{eq-relation2}
\frac{E^d_{final}}{E^e_{\Delta_c}}=\frac{1}{3+\alpha}\left[(2+\alpha)^\frac{3+\alpha}{1+\alpha} -1\right]\;.
\end{equation}
We can get the last relation Eqn.\ (\ref{eq-relation2}) by comparing 
expressions Eqn.\ (\ref{eq7d}) and Eqn.\ (\ref{eq6c}) directly.
These theoretical estimates are compared with numerical simulation results 
in Figures \ref{fig:final-d-vs-alpha}, \ref{fig:final-e-vs-alpha}. 
        
Now, if we put $\alpha=0$, we get these energy relations for uniform fiber 
threshold distribution:

\begin{equation}
\label{eq-relation1-uni}
\frac{E^d_{final}}{E^d_{\Delta_c}}=\left[(2)^3 -1\right]\;= 7.
\end{equation}
And
\begin{equation}
\label{eq-relation2-uni}
\frac{E^d_{final}}{E^e_{\Delta_c}}=\frac{1}{3}\left[(2)^3 -1\right]\;=\frac{7}{3}.
\end{equation}

That means the energy release during final catastrophic avalanche is  
much bigger than the  
the elastic energy stored in the 
system just before failure (final stable state when $\Delta=\Delta_c$).

It is commonly believed that during catastrophic events like earthquakes, 
landslides, dam collapses etc., the accumulated elastic energy
releases through avalanches \cite{cb97,brc15}. We observe a different scenario 
in this simple 
fiber bundle model where the system is doing  {\it work} during the 
catastrophic failure phase as the external force is still acting on the bundle.
As a result, the energy release (during catastrophic failure event) becomes 
much bigger than the elastic energy 
stored at the final stable phase.

\section{Energy-analaysis for Weibull distribution of thresholds}
We now consider the Weibull 
distribution, which has been used widely in material science \cite{hhp15}. The cumulative
Weibull distribution has a form:
 
\begin{equation}
\label{eq-Weibull-cumm}
P(x)= 1-\exp(-x^k),
\end{equation}
where $k$ is the Weibull index. 
Therefore the probability density takes the form:
\begin{equation}
\label{eq-Weibull-prob}
p(x)=kx^{k-1}\exp(-x^k).
\end{equation}
As the force has a maximum at the failure point $\Delta_c$, inserting $P(x)$ and $p(x)$ values in   
expression (Eqn.\ \ref{eq13})
we get
\begin{equation}
\label{eq-Weibull-Delta_c1}
\exp(-\Delta_c^k) - \Delta_c k\Delta_c^{k-1}\exp(-\Delta_c^k) = 0.
\end{equation}
 
From the above equation we can easily calculate the critical extension value 
as 
\begin{equation}
\label{eq-Weibull-Delta_c2}
\Delta_c =k^{-1/k}.
\end{equation}
The elastic energy at the critical extension $\Delta_c$ is: 
\begin{eqnarray}
\label{eq:Weibull-elastic}
E^{e}(\Delta_c)&=&\frac{N\kappa}{2}\ \Delta_c^2\left(1-P(\Delta_c)\right)\nonumber\\
&=&\frac{N\kappa}{2}\ \Delta_c^2\exp(-\Delta_c^k),\;
\end{eqnarray}
and the damage energy is
\begin{eqnarray}
\label{eq:Weibull-damage1}
E^d(\Delta_c)&=&\frac{N\kappa}{2}\ \int_0^{\Delta_c}\ d\delta \left[p(\delta)\delta^2\right]\nonumber\\
&=&\frac{N\kappa}{2}\ \int_0^{\Delta_c}\ kd\delta \left[\exp(-\delta^k)\delta^{k+1}\right].\;
\end{eqnarray}
Putting
\begin{equation}
\delta^k=u,
\end{equation}
we get
\begin{equation}
\label{eq:Weibull-damage2}
E^d(\Delta_c)=\frac{N\kappa}{2}\ \int_0^{\Delta_c^k}\ du \left[\exp(-u)u^{2/k}\right].\;
\end{equation}
This integral is exactly calculable for $k=1$ and $k=2$.
\subsection{Weibull distribution with $k=1$}
\begin{figure}[t]
\begin{center}
\includegraphics[width=8cm]{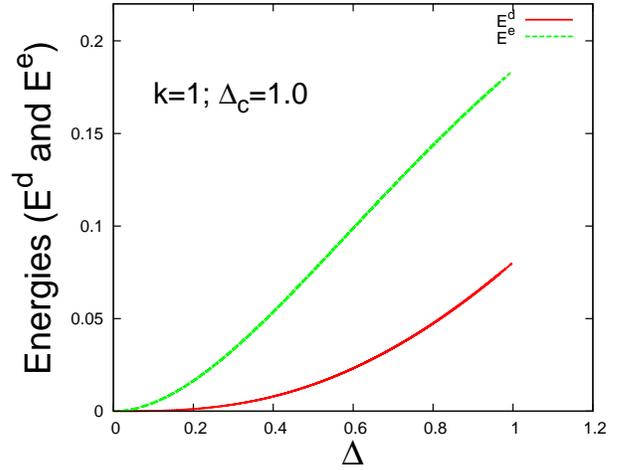}
\caption {\label{fig:Energies-vs-delta-Weibull-k1}
Damage energy and elastic energy {\it vs.\/} extension $\Delta$ (up to the 
failure point $\Delta_c$) for 
Weibull distribution of thresholds with Weibull index $k=1$. 
The simulation data (solid lines) are for a  
single bundle with $N=10^{7}$ fibers.}
\end{center}
\end{figure}
For Weibull index $k=1$, $\Delta_c=1$ and the damage energy expression at the 
failure point takes the form 
\begin{equation}
\label{eq:Weibull-damage-k1}
E^d(\Delta_c)=\frac{N\kappa}{2}\ \int_0^{1}\ du \exp(-u)u^2.\;
\end{equation}
Using integration by parts we arrive at the result
\begin{equation}
\label{eq:Weibull-damage-k1b}
E^d(\Delta_c=1)=\frac{N\kappa}{2}\ \left(2-5e^{-1}\right).\;
\end{equation}
We get the elastic energy at the failure point directly by putting $k=1$  in 
Eqn. (\ref{eq:Weibull-elastic}),
\begin{equation}
\label{eq:Weibull-elastic-k1}
E^{e}(\Delta_c=1)=\frac{N\kappa}{2e}\\;
\end{equation}
Therefore, the ratio between damage and elastic energies at the failure point
for Weibull distribution with $k=1$ is
\begin{equation}
\label{eq:energy_ratio-Weibull-k1}
\frac{E^d(\Delta_c)}{E^e(\Delta_c)}=2e -5 \;.
\end{equation}
In Figure \ref{fig:Energies-vs-delta-Weibull-k1}, we have shown numerical 
results of the variation 
of elastic and damage energies with strain for Weibull distribution 
(with $k=1$). The theoretical estimates of the ratio between damage and elastic
 energies at the failure point are compared with numerical results in Figure 
\ref{fig:ratio-vs-delta-Weibull}.  
\subsection{Weibull distribution with $k=2$}
\begin{figure}[t]
\begin{center}
\includegraphics[width=8cm]{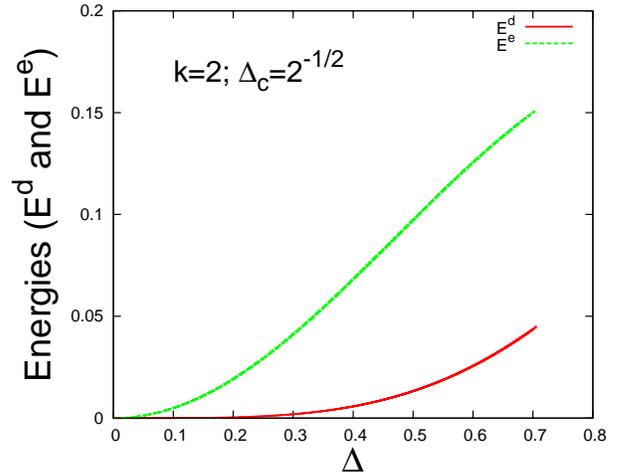}
\caption {\label{fig:Energies-vs-delta-Weibull-k2}
Damage energy and elastic energy {\it vs.\/} extension $\Delta$ (up to the 
failure point $\Delta_c$) for 
Weibull distribution of thresholds with Weibull index $k=2$. 
The simulation data (solid lines) are for a  
single bundle with $N=10^{7}$ fibers.}
\end{center}
\end{figure}
For Weibull index $k=2$, $\Delta_c=1/\sqrt{2}$ and the damage energy expression at the 
failure point is 
\begin{equation}
\label{eq:Weibull-damage-k2}
E^d(\Delta_c)=\frac{N\kappa}{2}\ \int_0^{1/2}\ du \exp(-u)u.\;
\end{equation}
Again, using integration by parts we arrive at the result
\begin{equation}
\label{eq:Weibull-damage-k2b}
E^d(\Delta_c=1/\sqrt{2})=\frac{N\kappa}{2}\ \left(1-\frac{3}{2 \sqrt{e}}\right).\;
\end{equation}
We get the elastic energy at the failure point directly by putting $k=2$  in 
Eqn.\ (\ref{eq:Weibull-elastic}):
\begin{equation}
\label{eq:Weibull-elastic-k2}
E^{e}(\Delta_c=1)=\frac{N\kappa}{2}\ \frac{1}{2\sqrt{e}}.\;
\end{equation}
Therefore, the ratio between damage and elastic energies at the failure point
for Weibull distribution with $k=2$ is
\begin{equation}
\label{eq:energy_ratio-Weibull-k2}
\frac{E^d(\Delta_c)}{E^e(\Delta_c)}=2\left(\sqrt{e} -\frac{3}{2}\right) \;.
\end{equation}
\begin{figure}[t]
\begin{center}
\includegraphics[width=8cm]{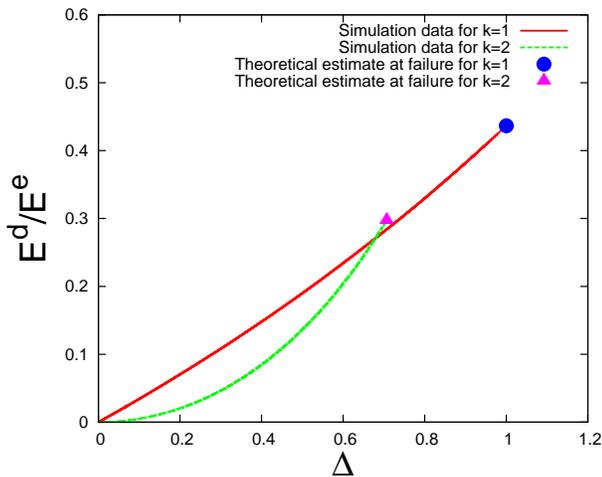}
\caption {\label{fig:ratio-vs-delta-Weibull}
Ratio between damage energy and elastic energy {\it vs.\/} extension $\Delta$ (up to the 
failure point $\Delta_c$) for a fiber 
bundle with Weibull distribution of thresholds. In simulation, we used a 
single bundle with $N=10^{7}$ fibers. Circle and triangle are the theoretical 
estimates 
(Eqns. (\ref{eq:energy_ratio-Weibull-k1}) and (\ref{eq:energy_ratio-Weibull-k2})) 
for the ratios at the failure point $\Delta_c$. }
\end{center}
\end{figure}
In Figure \ref{fig:Energies-vs-delta-Weibull-k2}, we have shown numerical 
results of the variation 
of elastic and damage energies with strain for Weibull distribution 
(with $k=2$). The theoretical estimates of the ratio between damage and elastic
 energies at the failure point is compared with numerical results in figure 
\ref{fig:ratio-vs-delta-Weibull}. In Appendix A, we give a general argument 
that elastic energy will be alaways bigger than damage energy at the 
critical (failure) point.  

\section{Elastic energy maximum}
\begin{figure}[t]
\begin{center}
\includegraphics[width=8cm]{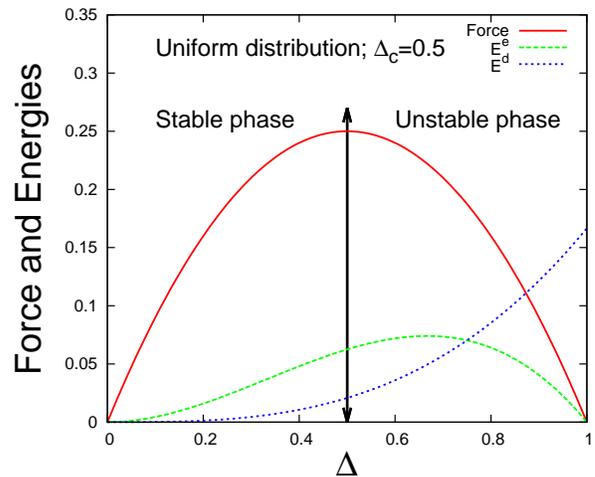}
\caption {\label{fig:phase-uniform}
Force, elastic energy and damage energy {\it vs.\/} extension $\Delta$ for a fiber 
bundle with uniform distribution of thresholds.}
\end{center}
\end{figure}

There are two distinct phases of the system: A stable phase for 
$0<\Delta\le\Delta_c$ and an unstable phase for $\Delta > \Delta_c$.
If we plot the elastic energy and damage energy  {\it vs.\/} $\Delta$, we 
see (Figure \ref{fig:phase-uniform}) that damage energy always increases with $\Delta$ but
elastic energy has a maximum at a particular value of $\Delta$, let us 
call it $\Delta_m$.
Can we calculate the exact value of $\Delta_m$ for a given threshold 
distribution? Is it somehow connected to $\Delta_c$?
In this section we are going to answer these questions.

\begin{figure}[t]
\begin{center}
\includegraphics[width=8cm]{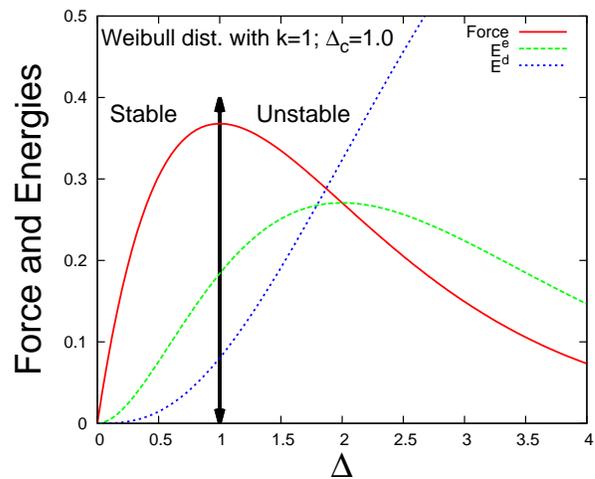}
\caption {\label{fig:phase-Weibull}
Force, elastic energy and damage energy {\it vs.\/} extension $\Delta$ for a fiber 
bundle with Weibull distribution of thresholds with $k=1$.}
\end{center}
\end{figure}
We recall the elastic energy expression Eqn.\ (\ref{eq6b}). 
If we differentiate the elastic energy with respect to the extension $\Delta$, 
we get   
\begin{equation}
\label{eq-en-diff}
\frac{dE^e(\Delta)}{d\Delta}=\frac{N\kappa}{2}\left[ 2\Delta\left(1-P(\Delta)\right)-\Delta^2 p(\Delta)\right],
\end{equation}
Which is $0$ at $\Delta_m$, with 
\begin{equation}
\label{eq-delm}
\Delta_m=\frac{2(1-P(\Delta_m))}{p(\Delta_m)}.
\end{equation}
If we consider a general power law type distribution $p(x)=(1+\alpha)x^\alpha$, within ($0,1$), we can write
\begin{equation}
\label{eq-delm-power}
\Delta_m=\left[\frac{2}{3+\alpha}\right]^\frac{1}{1+\alpha}=\Delta_c\left[\frac{2(2+\alpha)}{3+\alpha}\right]^\frac{1}{1+\alpha} > \Delta_c.
\end{equation}
For Weibull distribution $P(x)=1-\exp(-x^k)$, we can write
\begin{equation}
\label{eq-delm-Weibull}
\Delta_m=\left[\frac{2}{k}\right]^\frac{1}{k}=\Delta_c 2^\frac{1}{k} > \Delta_c.
\end{equation}
Therefore we can conclude that 
$\Delta_m$ is bigger than $\Delta_c$, i.e., elastic energy 
shows a maximum in the unstable phase (Figures \ref{fig:phase-uniform} and 
\ref{fig:phase-Weibull}).    
A more general treatment for the relation between $\Delta_m$ and $\Delta_c$ 
is given in the Appendix B.  

\section{Elastic energy inflection point: The warning sign of catastrophic 
failure}
Are there any prior indications of the catastrophic failure (complete 
failure) of a bundle under stress?
In the fiber bundle model, although the elastic energy has a maximum,
 it appears after the critical extenstion value, i.e., in the 
unstable phase of the system.  Therefore it can not help us to predict 
the catastrophic failure point of the system.

However, if we plot $dE^e/d\Delta$, the change of elastic energy with the 
change of extension value $\Delta$, we see that $dE^e/d\Delta$ has 
a maximum and, most importantly, this maximum appears before the 
critical extension value $\Delta_c$ (Figures \ref{fig:stable-unstable-uniform-data} and \ref{fig:stable-unstable-Weibull-data}). 
In this section we calculate the particular value of $\Delta$ 
at which $dE^e/d\Delta$ has a maximum. Let us call this $\Delta$ value 
$\Delta_{max}$.  We will also see whether there is a relation between 
$\Delta_{max}$ and $\Delta_c$.

\subsection{Theoretical analysis}
We recall the  expression for the derivative of elastic energy with respect to 
strain of extension Eqn.\ (\ref{eq-en-diff}). 
Taking derivative of the equation,
we get
\begin{equation}
\label{eq-en-d2}
\frac{d^2E^e(\Delta)}{d\Delta^2}=\frac{N\kappa}{2}\left[2\left(1-P(\Delta)\right)-4\Delta p(\Delta)-\Delta^2p'(\Delta)\right].
\end{equation}
Setting $d^2E^e(\Delta)/d\Delta^2=0$ at $\Delta=\Delta_{max}$
  we get for a general power law type distribution
\begin{equation}
\label{eq-delmax-power}
\Delta_{max}=\left[\frac{2}{(3+\alpha)(2+\alpha)}\right]^\frac{1}{1+\alpha}=\Delta_c\left[\frac{2}{3+\alpha}\right]^\frac{1}{1+\alpha}.
\end{equation}
This expression confirms that $\Delta_{max} < \Delta_c$ for $\alpha \ge 0$. 
For a Weibull distribution with index $k$, we can write
\begin{equation}
\label{eq-en-d2-Weibull}
\frac{d^2E^e(\Delta)}{d\Delta^2}=\frac{N\kappa}{2}\left[k^2\Delta^{2k}-(k^2+3k)\Delta^k+2\right]\exp(-\Delta^k).
\end{equation}
The solution  (of $d^2E^e(\Delta)/d\Delta^2=0$) with $(-)$ sign is the 
acceptable solution for the maximum. 
 Hence,   
\begin{eqnarray}
\Delta_{max}& =& \left[ \frac{(k+3)-\sqrt{(k+3)^2-8}}{2k} \right]^\frac{1}{k}\nonumber\\ 
&=& \Delta_c \left[ \frac{(k+3)-\sqrt{(k+3)^2-8}}{2} \right]^\frac{1}{k}\nonumber\\ 
&<& \Delta_c,
\end{eqnarray}
since

\begin{equation}
\left[ \frac{(k+3)-\sqrt{(k+3)^2-8}}{2} \right]^{1/k} < 1 \mathrm{\ } \forall \mathrm{\ } k > 0.
\end{equation}

A more general argument  is given in Appendix C for the relation between 
$\Delta_{max}$ and $\Delta_c$. 

\subsection{Comparison with simulation data}
\begin{figure}
\begin{center}
\includegraphics[width=8cm]{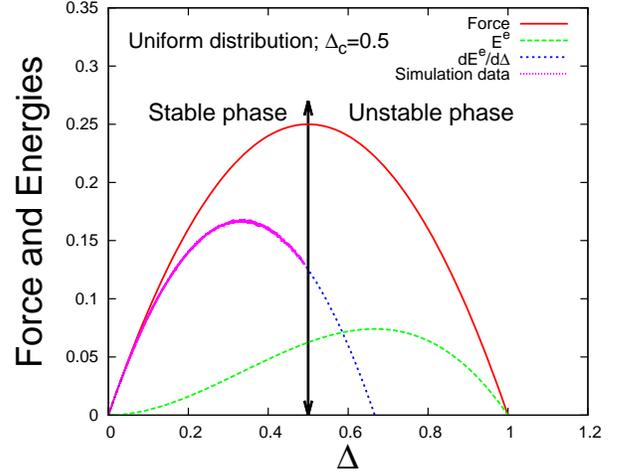}
\caption {\label{fig:stable-unstable-uniform-data}
Force and elastic energies {\it vs.\/} extension $\Delta$ for fiber 
bundles with uniform distribution of thresholds: Comparison with simulation data for a single bundle with $N=10^7$ fibers.}
\end{center}
\end{figure}

\begin{figure}
\begin{center}
\includegraphics[width=8cm]{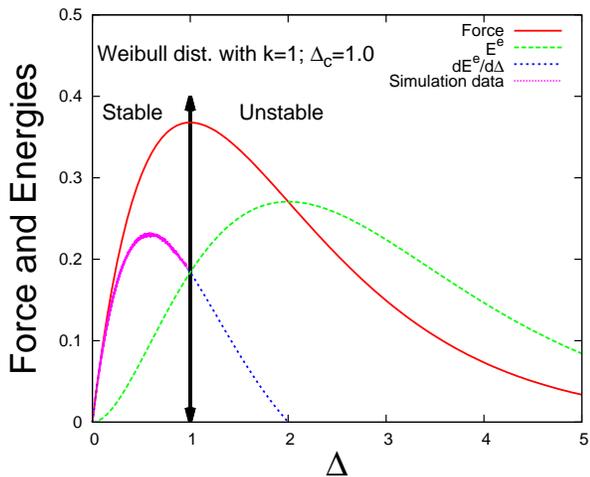}
\caption {\label{fig:stable-unstable-Weibull-data}
Force and elastic energies {\it vs.\/} extension $\Delta$ for fiber 
bundles with Weibull distribution of thresholds: Comparison with simulation data for a single bundle with $N=10^7$ fibers.}
\end{center}
\end{figure}

In Figures \ref{fig:stable-unstable-uniform-data} and 
\ref{fig:stable-unstable-Weibull-data} we compare the simulation results with 
the theoretical estimates. The simulations are done for a single bundle with 
large number  ($N=10^7$) of fibers and the agreement is convincing. 
We have used Monte Carlo technique to generate uncorrelated fiber thresholds 
that follow a particular statistical distributions (uniform and Weibull 
distributions). 
It is obvious 
that in simulations we can  measure energy values in the stable phase only.    
  
\subsection{Simulation results for a mixed threshold distribution}
\begin{figure}[t]
\begin{center}
\includegraphics[width=8cm]{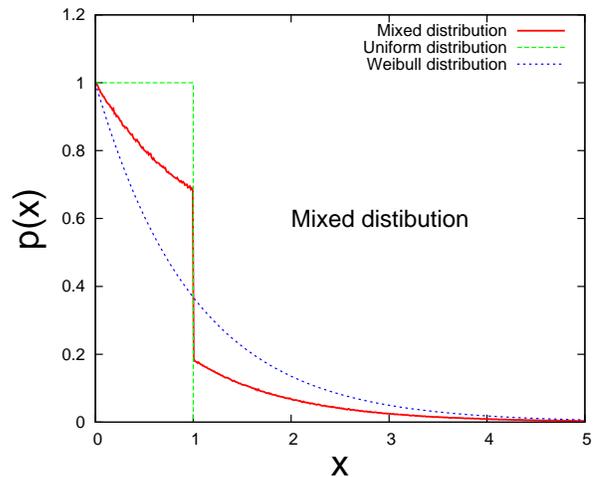}
\caption {\label{fig:mixed-distribution}
A mixture of uniform and Weibull ($k=1$) distributions.}
\end{center}
\end{figure}

\begin{figure}[t]
\begin{center}
\includegraphics[width=8cm]{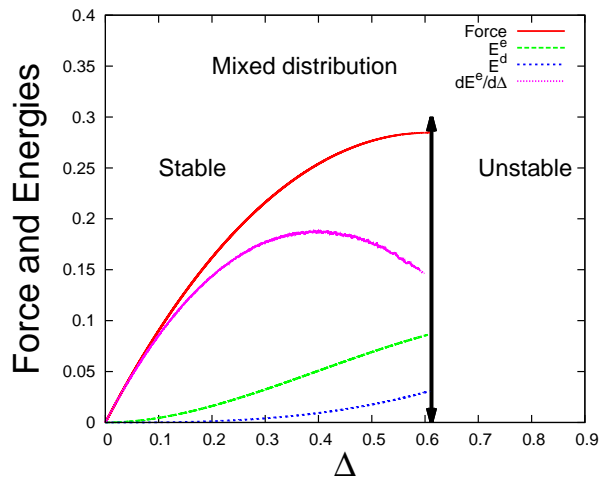}
\caption {\label{fig:mixed-energies}
Force and elastic energies {\it vs.\/} extension $\Delta$ for a fiber 
bundle ($10^7$ fibers) with a mixed distribution of thresholds.}
\end{center}
\end{figure}

Now we choose a mixed fiber 
threshold distribution, i.e., a threshold distribution.  
Can we see similar signature (maximum of 
$dE^e/d\Delta$ appears before $\Delta_c$) as we have seen in previous section?
The chosen distribution is  a mixture of uniform distribution and Weibull 
distribution 
($k=1$) which is shown in Figure \ref{fig:mixed-distribution}. 
We assign strength thresholds to $N/2$ fibers from 
a uniform distribution and to $N/2$ fibers from  
a Weibull distribution.  
The simulation result (figure \ref{fig:mixed-energies}) reveals that 
$dE^e/d\Delta$ has a maximum which 
appears before the failure point $\Delta_c$ and $\Delta_{max}$ 
is 
somewhere in between the respective $\Delta_{max}$ values for uniform and 
Weibull 
threshold distributions --- as expected intuitively.         
\section{Discussions}

The Fiber Bundle Model has been used as a standard model for studying 
stress-induced fracturing in
composite materials.  In the Equal-Load-Sharing version of the model, 
all intact fibers share the load equally.  
In this work we have chosen the ELS models and we have studied the energy 
budget of the model for the entire failure process, starting from intact 
bundle up to the catastrophic failure point where the bundle collapses 
completely.  
Following the standard definition of elastic and damage 
energies from continuous damage mechanics framework, we have calculated the 
energy relations at the failure points for different types of fiber threshold 
distributions (power law type and Weibull type).  
At the critical or catastrophic failure point,
the elastic energy is always larger than the total damage energy. 
Another important observation is that
the elastic energy variation
has a distinct peak before the catastrophic failure point. Also, 
the energy-release
during final catastrophic event is much bigger than the elastic energy stored
in the system at the
failure point (see section III and section IV). 
Our simulation results on a single bundle with large numbers ($10^7$)  of 
fibers, show perfect agreement with the
theoretical estimates.  We have chosen a {\it{single bundle}}, keeping in 
mind that for prediction purposes it is important and 
necessary that the warning sign can be seen in a single sample. 

These
observations can form the basis of a prediction scheme by finding the 
correlation between the position
(strain or stretch level) of elastic energy variation peak and the actual 
failure point. Moreover, it is
also possible to predict the size (energy release) of the final catastrophic event 
by measuring the stored
elastic energy of the system at the failure point.

Our observations in this work have already opened up some scientific questions 
and challenges: what happenes for Local-Load-Sharing (LLS) models \cite{hp91}? 
Does the elastic energy variation show similar peaks before the catastrophic 
failure point? Can we measure and analyze the elastic energy during a 
rock-fracturing test \cite{ps15} in terms of the applied strain?   
  
Our next article will resolve some of these issues --we are now working on 
energy budget of LLS models.

\begin{acknowledgments}
This work was partly supported by the
Research Council of Norway through its Centers of Excellence funding
scheme, project number 262644.
\end{acknowledgments}
\section{Appendix}
As stated in section II, the elastic energy in the system at extension $\Delta$ is

\begin{equation}
E^{e}(\Delta) = \frac{N\kappa}{2} \Delta^{2} (1 - P(\Delta)),
\label{eq:elastic_energy}
\end{equation}
where $P(\Delta)$ is the cumulative probability distribution of the fiber thresholds. The force per fiber $\sigma = F/N$ required to continue the breaking process at a given extension $\Delta$ is

\begin{equation}
\sigma(\Delta) = \kappa \Delta (1-P(\Delta))
\label{eq:sigma}
\end{equation}
The critical extension $\Delta_{c}$ where the bundle collapses is hence given by

\begin{equation}
0 = \left. \frac{d\sigma}{d\Delta} \right|_{\Delta_{c}} = \kappa (1 - P(\Delta_{c}) - \Delta_{c} p(\Delta_c)).
\label{eq:d_sigma}
\end{equation}

\subsection{Elastic versus damage energy at the critical point}
Numerical data seems to suggest that $E^{e} > E^{d}$ at the critical point for most threshold distributions. Let us try to prove this analytically by investigating the difference between elastic and damage energy:

\begin{equation}
\begin{aligned}
E_{diff}(\Delta)& = E^e(\Delta) - E^d(\Delta)\\
& = \frac{N\kappa}{2} \left[ \Delta^2 (1 - P(\Delta)) - \int_{0}^{\Delta} dx \text{ } x^{2} p(x) \right].\\
\label{eq:diff_energy}
\end{aligned}
\end{equation}
The derivative of this energy difference is

\begin{equation}
\begin{aligned}
\frac{dE_{diff}}{d\Delta} &= \frac{N\kappa}{2} \left[ 2\Delta (1 - P(\Delta)) - \Delta^2 p(\Delta) - \Delta^2 p(\Delta) \right] \\
&= N\kappa\Delta \left[ 1 - P(\Delta) - \Delta p(\Delta) \right] \\
&= N\Delta \frac{d\sigma}{d\Delta}.
\end{aligned}
\end{equation}
We can now express the energy difference in terms of the forces acting on the fiber bundle. We integrate this expression to find

\begin{equation}
\begin{aligned}
E_{diff}(\Delta) &= \int_{0}^{\Delta} dx \text{ } \frac{dE_{diff}}{dx} = N \int_{0}^{\Delta} dx \text{ } x \frac{d\sigma}{dx} \\
&= N \left[ \Delta \sigma(\Delta) - \int_{0}^{\Delta} dx \text{ } \sigma(x) \right]
\end{aligned}
\end{equation}
by partial integration. In particular, this gives the result

\begin{equation}
E_{diff}(\Delta_{c}) = N \left[ \Delta_{c} \sigma_{c} - \int_{0}^{\Delta_{c}} 
 dx \sigma(x) \right]
\end{equation}
at the critical point. Since $\sigma_{c} = \text{max } \sigma(\Delta)$, we see that $E_{diff}(\Delta_{c}) > 0$ for all threshold distributions. The only exception possible is a threshold distribution with a constant force $\sigma(\Delta) = \sigma_{c}$. But this results in a lower cut-off $\Delta_{0} = \Delta_{c} > 0$ (for the threshold distribution to be normalizable), and then $E^{e}(\Delta_{c}) > E^{d}(\Delta_{c}) = 0$.
\subsection{Elastic energy maximum point}
First, rewrite Eqn. (\ref{eq:d_sigma}) as

\begin{equation}
1 = g(\Delta_{c}) \equiv \frac{1-P(\Delta_{c})}{\Delta_{c} p(\Delta_{c})}.
\label{eq:d_c}
\end{equation}
This definition of $g(\Delta)$ will be useful in the following derivations. The maximum of the elastic energy is found at an extension $\Delta_m$, which is given by

\begin{equation}
0 = \left. \frac{dE^{e}}{d\Delta} \right|_{\Delta_{m}} \propto 2\Delta_{m} (1 - P(\Delta_{m})) - \Delta_{m}^{2} p(\Delta_{m}),
\end{equation}
i.e.,

\begin{equation}
\frac{1}{2} = \frac{1-P(\Delta_{m})}{\Delta_{m} p(\Delta_{m})} = g(\Delta_{m}).
\label{eq:d_m}
\end{equation}
Comparing this expression to Eqn. (\ref{eq:d_c}) allows us to find a relation between $\Delta_{c}$ and $\Delta_{m}$. We investigate the function $g(\Delta)$:

\begin{equation}
\begin{aligned}
g(\Delta) = \frac{1-P(\Delta)}{\Delta p(\Delta)} &= \frac{1-P(\Delta)}{1-P(\Delta)-\frac{d\sigma}{d\Delta}} \\
&=
\begin{cases}
        < 1 & \text{ for\ } \frac{d\sigma}{d\Delta} < 0 \\
        > 1 & \text{ for\ } \frac{d\sigma}{d\Delta} > 0
\end{cases}.
\end{aligned}
\label{eq:def_g}
\end{equation}

It is clear from Eqn. (\ref{eq:def_g}) that for a threshold distribution with only a single maximum in the load curve, $g(\Delta_{m}) = 1/2$ must occur in the unstable phase, i.e., $\Delta_{m} > \Delta_{c}$.

\subsection{Elastic energy inflection point}
The elastic energy maximum occurs after the critical point and is hence unsuitable as a predictor for failure. But what about the maximum of the \textit{derivative} of the elastic energy, the inflection point $\Delta_{max}$? As stated in the section VI:

\begin{equation}
\frac{d^2E^{e}}{d\Delta^2} \propto 2(1-P(\Delta)) - 4\Delta p(\Delta) - \Delta^2 p'(\Delta).
\end{equation}
Setting this second derivative to zero and rearranging terms gives the equation

\begin{equation}
2(1-P(\Delta_{max})) - 4\Delta_{max} p(\Delta_{max}) = \Delta_{max}^2 p'(\Delta_{max}).
\end{equation}
To find a relation between $\Delta_{max}$ and $\Delta_{c}$ we substitute $\Delta_{max} p'(\Delta_{max}) = - 2p(\Delta_{max}) - \left. d^{2}\sigma/d\Delta^{2} \right|_{\Delta_{max}}$ and then divide by $2\Delta_{max} p(\Delta_{max})$. After rearranging terms the result is

\begin{equation}
1 = g(\Delta_{max}) + \frac{\left. \frac{d^{2}\sigma}{d\Delta^{2}} \right|_{\Delta_{max}}}{2\Delta_{max} p(\Delta_{max})}.
\end{equation}
From this expression and Eqn. (\ref{eq:def_g}) we can see that at $\Delta_{max}$, $d\sigma/d\Delta$ and $d^{2}\sigma/d\Delta^{2}$ must have different signs.

Let's once again assume that we are working with a threshold distribution that has only a single maximum in its load curve. Then $d\sigma/d\Delta > 0$ corresponds to the stable phase and $d\sigma/d\Delta < 0$ corresponds to the unstable phase. We see that any threshold distribution with $d^{2}\sigma/d\Delta^{2} < 0$ everywhere in the unstable phase must have $\Delta_{max}$ in the stable phase, i.e., $\Delta_{max} < \Delta_{c}$.

This is a general (but weak) condition that is sufficient, but not necessary, for $\Delta_{max} < \Delta_{c}$.


\begin{thebibliography}{}
\bibitem{k96} D.\ Krajcinovic, {\it Damage mechanics\/} (Elsevier, Amsterdam, 1996).

\bibitem{hhlp08} M. Henkel, H. Hinrichsen, S. Lubeck and M. Pleimling, {\it Non-equilibrium phase transitions} vol. 1 (Springer, Berlin, 2008).

\bibitem{bb11} D.\ Bonamy and E.\ Bouchaud, {\it Failure of heterogeneous materials: A dynamic phase transition?}, 
Phys.\ Rep.\ {\bf 498}, 1 (2011).

\bibitem{a15} S.\ G.\ Abaimov, {\it Statistical physics of non-thermal
phase transitions\/} (Springer Verlag, Heidelberg, 2015).

\bibitem{b15} E.\ Berthier, {\it Quasi-brittle failure of
heterogeneous materials: damage statistics and localization,\/}
thesis, Universit{\'e} de Paris 6 (2015).

\bibitem{cb97} B. K. Chakrabarti and L. G. Benguigui, {\it Statistical Physics of Fracture and Breakdown in Disordered Solids}, Oxford University Press, Oxford (1997).
 
\bibitem{brc15} S.\ Biswas, P.\ Ray and B.\ K.\ Chakrabarti, {\it
Statistical physics of fracture, breakdown, and earthquake\/}
(Wiley-VCH, Berlin, 2015).

\bibitem{vm91} P. Villa and E. Mahieu, {\it Breakage patterns of human long bones}, J. Hum. Evol. {\bf 21} 1, 27-48 (1991).

\bibitem{e84} D. Etzion, {\it Moderating effect of social support on the stress-burnout relationship}, J. App. Psych., {\bf 69}(4), 615-622 (1984). 

\bibitem{rgrcg96} A. J. Ramirez, J. Graham, M. A. Richards, A. Cull and W. M. Gregory, {\it Mental health of hospital consultants: the effect of stress and satisfaction at work}, Lancet {\bf 347} 724-728 (1996). 

\bibitem{p26} F.\ T.\ Peirce, {\it Tensile tests for cottom yarns. ``The weakest link" theorems on the strength of long and composite specimens}, J.\ Text Ind., {\bf 17}, 355 (1926).

\bibitem{d45} H.\ E.\ Daniels, {\it The statistical theory of the strength of bundles of threads}, Proc.\ Roy. Soc.\ Ser.\ A {\bf 183} 243 (1945).

\bibitem{phc10} S.\ Pradhan, A.\ Hansen and B.\ K.\ Chakrabarti, {\it Failure processes in elastic fiber bundles}, 
Rev.\ Mod.\
Phys.\ {\bf 82}, 499 (2010).

\bibitem{hhp15} A.\ Hansen, P.\ C.\ Hemmer and S.\ Pradhan, {\it The
fiber bundle model: Modeling Failure in Materials\/} (Wiley-VCH, Berlin, 2015).

\bibitem{phr18} S. Pradhan, A. Hansen and P. Ray, {\it A Renormalization Group Procedure for Fiber Bundle Models}, 
Front. Phys. {\bf 6}, 65 (2018).

\bibitem{hp91} D.\ G.\ Harlow and S.\ L.\ Phoenix, {\it Approximations for the strength distribution and size effects in an idealized lattice model of material breakdown}, J.\ Mech.\ Phys.\
Sol.+ {\bf 39}, 173 (1991).

\bibitem{ps15}  S. Pradhan, A. Stroisz, E. Fjar, J. Stenebraten,  H. Lund, E. Sonstebo, {\it Stress-Induced 
Fracturing of Reservoir Rocks: Acoustic monitoring and mCT Image Analysis}, Rock Mech. and Rock Eng. 
Vol 48.(6) s. 2529-2540 (2015).

\end{thebibliography}
\end{document}